\documentclass[a4paper,12pt]{article}
\usepackage{mathrsfs}
\usepackage{graphicx} 
\usepackage{epstopdf}

\begin{document}

\hoffset = -1truecm \voffset = -2truecm \baselineskip = 10 mm

\title{\bf The gluon condensation in hadron collisions}

\author{
{\bf Wei Zhu$^a$\footnote{Corresponding author, E-mail:
wzhu@phy.ecnu.edu.cn}}, {\bf Qihui Chen$^b$}, {\bf Zhiyi Cui$^a$} and {\bf Jianhong Ruan$^a$}
\\
\normalsize $^a$Department of Physics, East China Normal University,
Shanghai 200241, P.R. China \\
\normalsize $^b$School of Physical Science and Technology Southwest Jiaotong University,\\
Chengdu 610031, P.R. China\\
}

\date{}

\newpage

\maketitle
\begin{abstract}

    Gluons may converge to a stable state at a
critical momentum in hadrons. This gluon condensation
is predicted by a nonlinear QCD evolution equation. We review the understanding
of the gluon condensation and present a clear physical picture that produces the gluon condensation from
the colour glass condensate. We summarize the applications of the GC effect in the $p-p(A)$ collisions and
predict that the $p-Pb$ and $Pb-Pb$ collisions at the LHC are close to
the energy region of the gluon condensation.
We warn that for the next generation of hadron colliders with the increasing of the collision energy,
the extremely strong gamma-rays will be emitted in a narrow space
of the accelerator due to the gluon condensation effect. Such artificial mini gamma-ray bursts
in the laboratory may damage the detectors.

\end{abstract}

{\bf keywords}:  Gluon condensation; Antishadowing effect; Next generation of LHC

\vskip 1truecm

\newpage
\begin{center}
\section{Introduction}
\end{center}

    Hadron in infinite momentum frame is consisted of partons (i.e., quarks and gluons).
Gluons distribute mainly in the small $x$ range ($x$ is the Bjorken variable) and they dominate the high energy hadronic processes.
The parton distribution functions are evolved according to the evolution equations based on perturbative QCD (pQCD) in the standard model. For example, the gluon distribution functions satisfy the Dokshitzer-Gribov-Lipatov-Altarelli-Parisi (DGLAP) equation in a broad kinematic range, its elemental amplitude is shown in Fig. 1a, where the correlations among initial partons are neglected [1]. Obviously, these correlations should be considered when the gluon density becomes large in very small $x$ region, where the wave functions of the initial partons begin to overlap.  Adding the initial gluons to the elemental amplitude of the DGLAP equation, one can get a series of modified elemental amplitudes in Figs. 1(b)-1(d), corresponding to the Balitsky-Fadin-Kuraev-Lipatov (BFKL) [2], Gribov-Levin-Ryskin-Mueller-Qiu
-Zhu-Ruan-Shen (GLR-MQ-ZRS) [3,4] and Zhu, Shen, Ruan (ZSR) [5] equations, respectively. For convenience, we use the names of authors
to mark the different evolution equations.
According to the standard quantum field theory, a complete evolution equation should include the contributions of all possible
Feynman diagrams for energy-momentum conservation and infrared (IR) safety. Works [4,5] used the time ordered perturbative theory
(TOPT) to derive the evolution equations in Fig. 1. Interestingly, one of the resulting evolution equation (ZSR)
predicts that the gluons in hadron may converge at a critical momentum
(Fig. 3). This is the gluon condensation (GC).

\begin{figure}
    \begin{center}
        \includegraphics[width=0.8\textwidth]{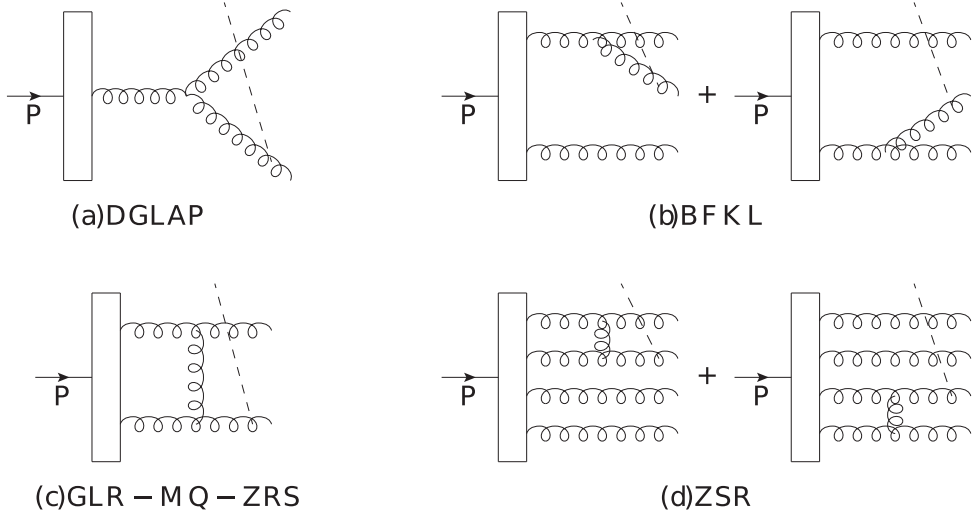}
        \caption{The corrections of the initial gluons to an elemental
amplitude of the DGLAP equation (a) [1] and they lead to (b) the BFKL
equation [2], (c) the GLR-MQ-ZRS equation [3-4] and (d) the ZSR evolution
equation [5], respectively.  The dashed line is a virtual current which
probes gluons.}\label{Fig.1}
    \end{center}
\end{figure}

    The GC is displayed in the numerical solution of an approximate evolution equation. A question is that the
GC does exactly exist in the nature? or is it just an approximate solution of the equation, once the approach is improved, will the GC
disappear? This work attempts to answer the above question in the following two steps.

(1) We briefly
review the GC-theory in Sec. 2, where we track the formation processes of the GC through Figs. 4 and 5.
We noticed that the transverse momenta of the splitting and fusing gluons in hadrons at high energy processes are randomly changed. It leads to the so-called Lipatov-singularity [2]. This infrared singularity should be regularized by summing all relating Feynman diagrams according to the standard quantum field theory, after which, we can obtain Eq. (2.1), where both the linear and nonlinear parts have the regularized Lipatov-singularities. The nonlinear part in Eq. (2.1) converts the weak jumps on the gluon distributions into the strong chaotic oscillations. The chaotic oscillation generates the strong antishadowing effect, and the later leads to the GC.
Thus, we show a more clear physical picture of the GC, which is quantitatively described
by a delta-like function with the undetermined GC-critic momentum $(x_c,k_c)$.

(2) Then we turned to discuss the applications of the GC effect in Sec. 3.
The GC should induce significant effects in
the proton collisions if gluons with the GC-critic momentum  enter
the measuring energy region. Unfortunately, the value of $(x_c,k_c)$
can not been entirely determined in the theory since it
relates to the unknown input conditions and high order modifications. We have not directly
observed the GC-effects at the
Large Hadron Collider (LHC).  Therefore, the GC effect was used to analyse
the astrophysical observations in [6-9], since the proton energy
may be accelerated to cause the GC-effect in the hadronic collisions.
Although the astrophysical phenomena relate to many complex factors, the GC effect as a general physical effect should be shown in
the astronomical observation phenomenon.
Specifically, cosmic gamma-rays can be generated in $p+p\rightarrow \pi^0\rightarrow 2\gamma$.
The sharp peak in the momentum distribution of gluons caused by the GC effect in Fig. 3 can suddenly increase
the cross section in the hadron-hadron
collisions, and results in a typical broken power law in the gamma-ray energy spectra, where the broken energy
$E_{\pi}^{GC}$ is determined by the critical momentum. After a brief summary we
discuss the GC-threshold in the LHC energy range.  We find that the interaction energies in the $p-Pb$ and $Pb-Pb$ collisions in LHC are close to the GC-energy region. Note that the cross section of pion production
in the hadronic colliders can be increased by several orders of magnitude due to the GC effect, and about half of the proton kinetic energy
converted to the photons of energy $E_{\gamma}=m_{\pi}/2$ with an extra intensity in a narrow space at the moment of collision. Therefore, we issue a warning for the next accelerator plans since such unexpected intense gamma rays in the laboratory may damage the detectors.
Finally, the discussions and summary are given in Sec. 4, where we explain why the Balitsky-Kovchegov (BK) [10] does not have the GC solution. We also show the self-consistence among the popular QCD evolution equations.

\newpage
\begin{center}
\section{The GC effect in the ZSR equation}
\end{center}

    The ZSR equation at the cylindrically symmetric approximation
reads [5]

$$-x\frac{\partial F(x,k_\perp ^2)}{\partial x}$$
$$=\frac{3\alpha_{s}k_\perp^2}{\pi}\int_{k^2_0}^{\infty} \frac{d k'^2_\perp}
{k'^2_\perp}\left\{\frac{F(x,k'^2_\perp)-F(x,k^2_\perp)}
{\vert
k'^2_\perp-k^2_\perp\vert}+\frac{F(x,k^2_\perp)}{\sqrt{k^4_\perp+4k'^4_\perp}}\right\}$$
$$-\frac{81}{16}\frac{\alpha_s^2}{\pi R^2_N}\int_{k^2_0}^{\infty}
\frac{d k'^2_\perp}{k'^2_\perp}\left\{\frac{k^2_\perp F^2(x,k'^2_\perp)-k'^2_\perp F^2(x,k^2_\perp)}
{k'^2_\perp\vert{
k'^2_\perp-k^2_\perp\vert}}+\frac{F^2(x,k^2_\perp)}{\sqrt{k^4_\perp+4k'^4_\perp}}\right\},
\eqno(2.1)$$where $F$ is the unintegrated gluon distribution and
the linear part describes the contributions of the BFKL evolution.
We used $F(x/2,k_\perp^2)\simeq F(x,k_\perp^2)$ near the
saturation scale; the value of $R_N=4~GeV^{-1}$ is fixed by fitting
the available experimental data of the proton structure functions.

    We take two saturated inputs to evolve Eq. (2.1) starting from $x_0=4\times 10^{-5}$.
They are the Golec-Biernat and Wusthoff (GBW)
model [11]

$$F_{GBW}(x_0,k_\perp^2)=\frac{3\sigma_0}{4\pi^2\overline{\alpha}_s}R^2_0(x)k_\perp^4
\exp(-R^2_0(x)k_\perp^2), \eqno(2.2)$$where $\sigma_0=29.12~mb$,
$R_0(x)=1/Q_s$, $Q_s=1~GeV$, $\overline{\alpha}_s=0.2$, and
the Kharzeev-Levin (KL) model [12]

$$F_{KL}(x_0,k_\perp^2)=\left\{\begin{array}{ll}
f_0 k_\perp^2 ~if~ k_\perp^2<Q^2_s\\
f_0Q^2_s ~if~ k_\perp^2> Q^2_s, \end{array}\right. \eqno(2.3)$$where $f_0=10$.
These two inputs are plotted in Fig.2, they describe the so called colour glass condensate (CGC) [13].
Note that in the calculations we take $F(x,k_\perp^2)=0$ if $F(x,k_\perp^2)<0$
since $F(x,k_\perp^2)\geq 0$ according to the definition of the gluon
distribution.

\begin{figure}
    \begin{center}
        \includegraphics[width=0.8\textwidth]{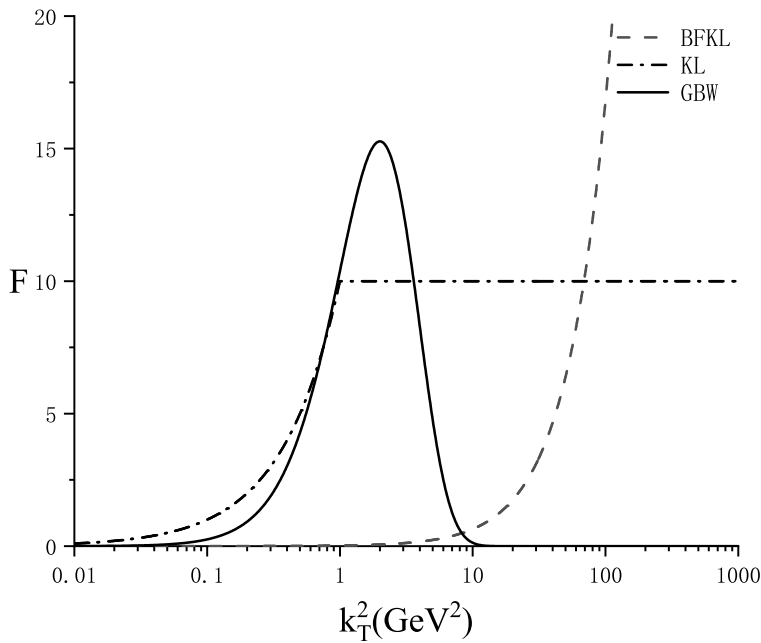}
        \caption{The GBW-, KL- and BFKL-inputs.}\label{Fig.2}
    \end{center}
\end{figure}

\begin{figure}
\begin{center}
    \includegraphics[width=0.8\textwidth]{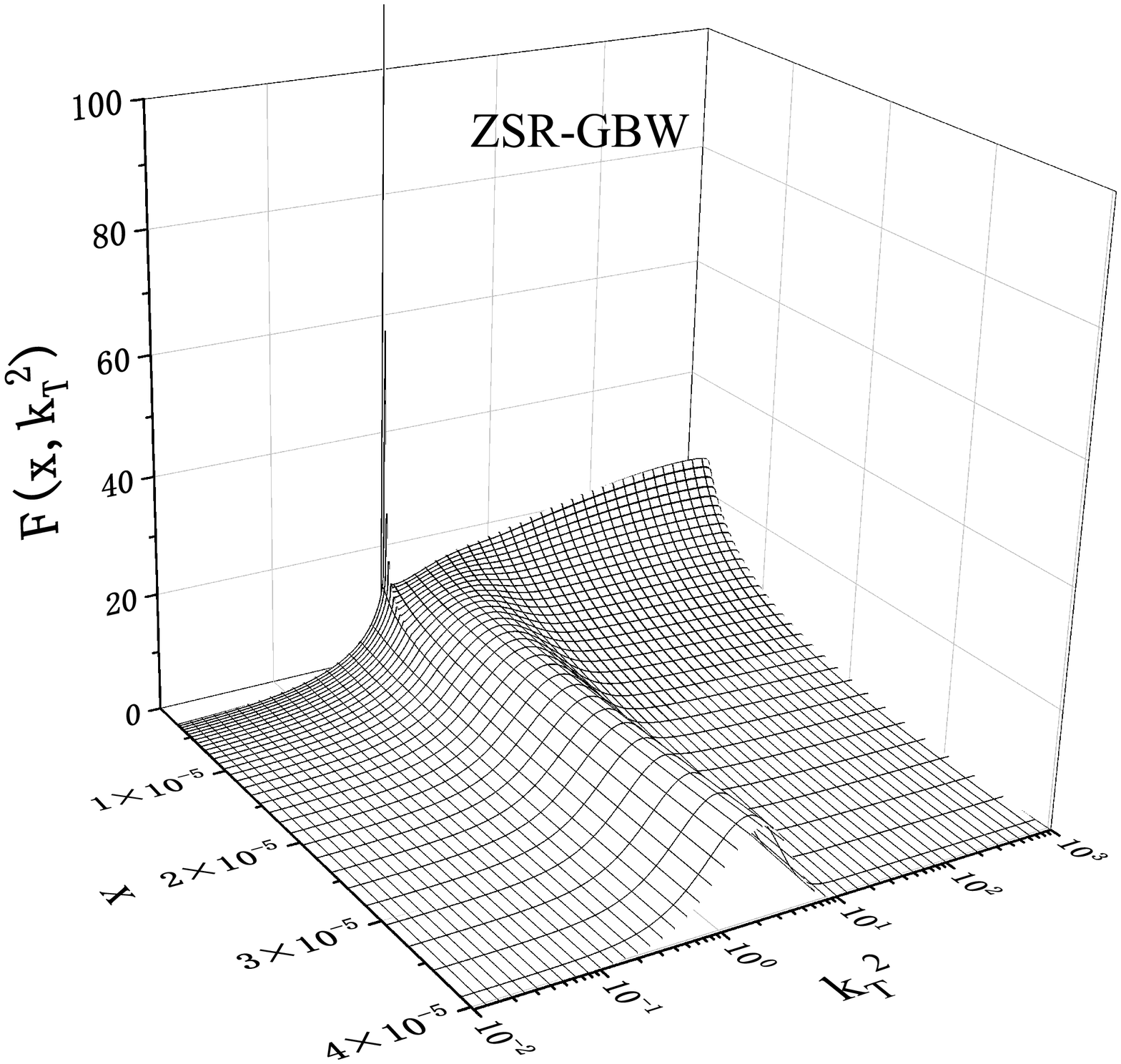} 
    \includegraphics[width=0.8\textwidth]{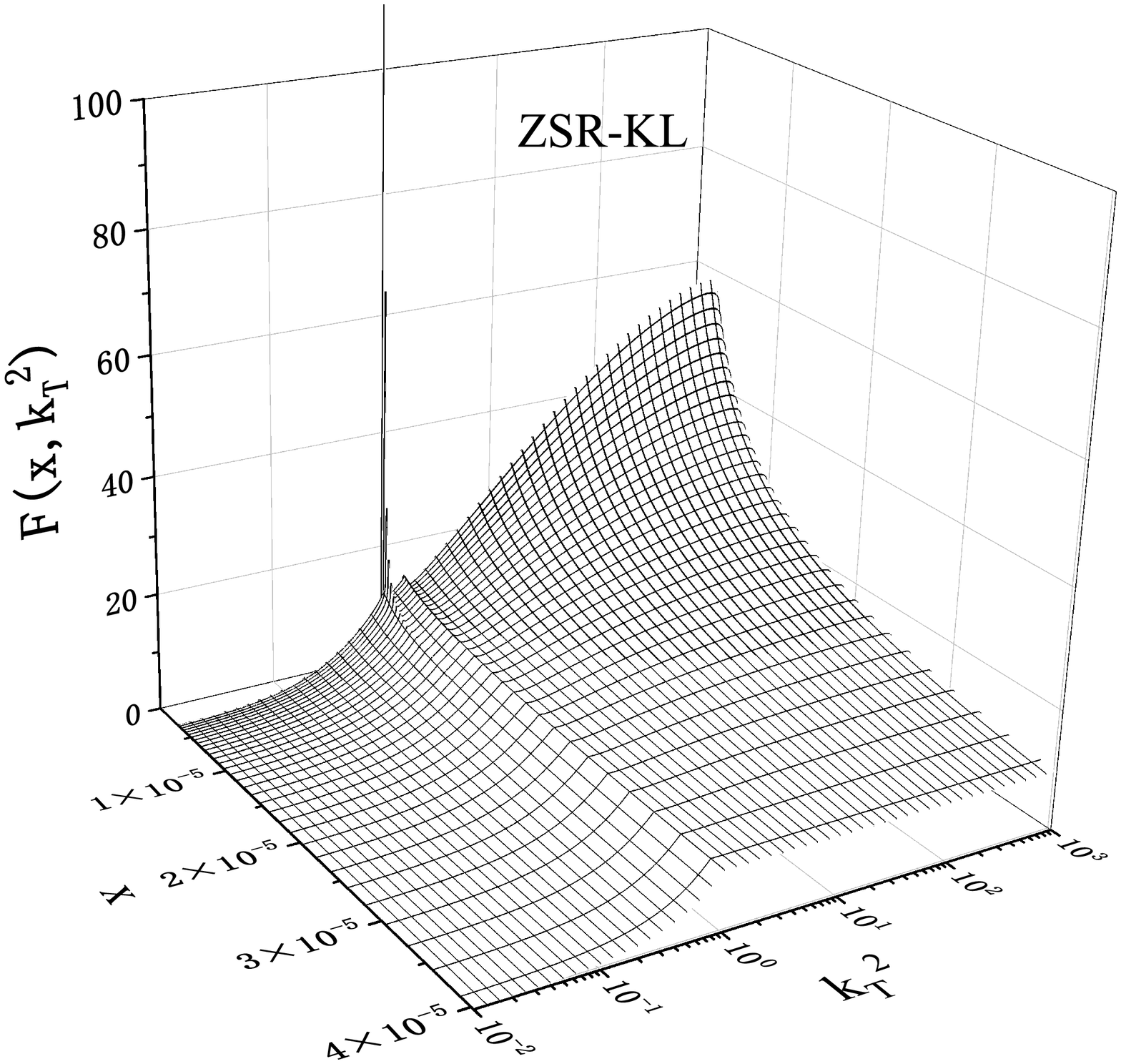} 
     \caption{The solutions of Eq. (2.1) with the GBW- and KL-inputs, where the Runge-Kutta method is used.}\label{Fig.3}
\end{center}
\end{figure}

    The solutions of $F(x,k^2_\perp)$ in three-dimensional representation are given in Fig. 3,
they show that gluons in the proton are converged
at a critical momentum $(x_c.k^2_c)$.  In order to expose its origin, we study the structure of Eq. (2.1).
An important character of the BFKL dynamics is that the tracks of gluon motion on the transverse momentum $k_\perp$-plane
are randomly distributed in the evolution. One can find that the variable $k'^2_\perp$ may cross over $k^2_\perp$ and generates the IR-divergences both in the linear and nonlinear kernels as shown in Eq. (2.1). However, any physical processes must be IR-safe. The most used nonlinear modifications to the BFKL equation take the way to avoid the IR-divergences, while these IR-divergences are canceled by using the TOPT cutting rule in works [4,5].

    We draw the integrated function $R(x_0,k^2_\perp, k'^2_\perp)\sim k'^2_\perp$
of Eq. (2.1) at four different values of $k^2_T$ using two inputs (2.2) and (2.3) in Fig. 4. For comparison, a similar figure
is also drawn but evolved by the linear BFKL equation with a normal BFKL-input [14]

$$F_{BFKL}(x_0,k_\perp^2)=\beta\sqrt{k^2_\perp}\frac{x_0^{-\lambda_{BFKL}}}{\sqrt{\ln(1/x_0)}}
\exp\left(-\frac{\ln^2(k^2_\perp/k^2_s)}
{2\lambda"\ln(1/x_0)}\right),\eqno(2.4)$$where $\lambda_{BFKL}=12\overline{\alpha}_s/(\pi\ln 2), \lambda"=32\overline{\alpha}_s,
\beta=0.01$ and $k^2_s=1~GeV^2$. One can find a common feature of the results: having jumps at $k'^2_\perp=k^2_\perp$. It is
the direct result of the Lipatov-singularity.
After integral, these jumps generate a series of weak perturbations to the gluon distribution $F(x, k^2_\perp)$.
The above perturbations are independent in the linear BFKL equation and their effects are
negligibly small. In this case the solutions are almost smooth curves
in the $k_\perp^2$-dependence. However, the special nonlinear terms in Eq. (2.1) may occur
the coupling among random perturbations and generates chaos.
A standard criterion of chaos is that the system has the positive
Lyapunov exponent, which indicates a strong sensitivity to small
changes in the initial conditions. The Lyapunov exponents of the gluon
distributions in Eq. (2.1) with the inputs Eqs. (2.2) and (2.3) are presented in
Fig. 5. The positive Lyapunov exponent of the solutions of Eq. (2.1) is a
strong evidence for the chaotic solution.

\begin{figure}
\begin{center}
    \includegraphics[width=0.8\textwidth]{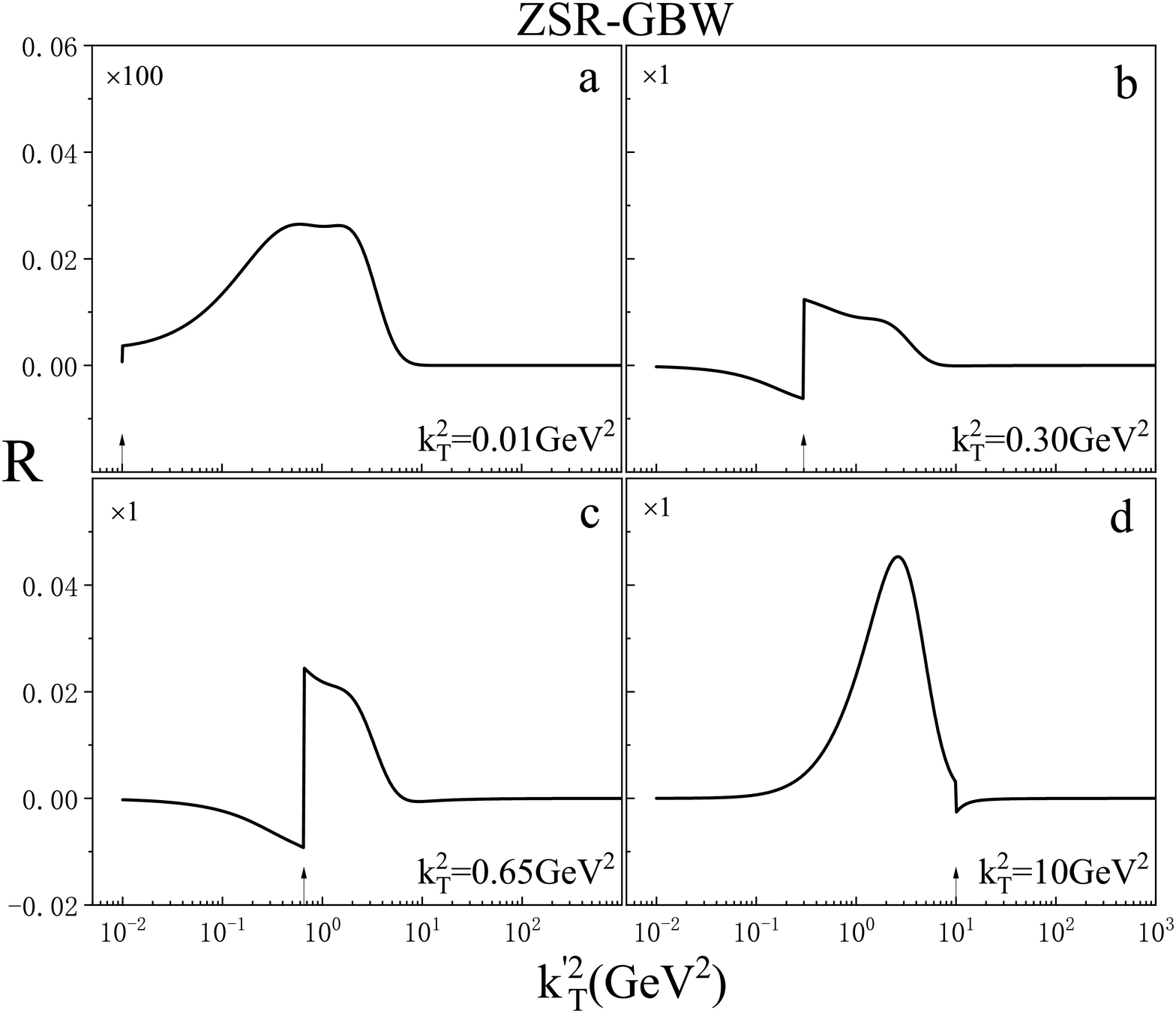} 
    \includegraphics[width=0.8\textwidth]{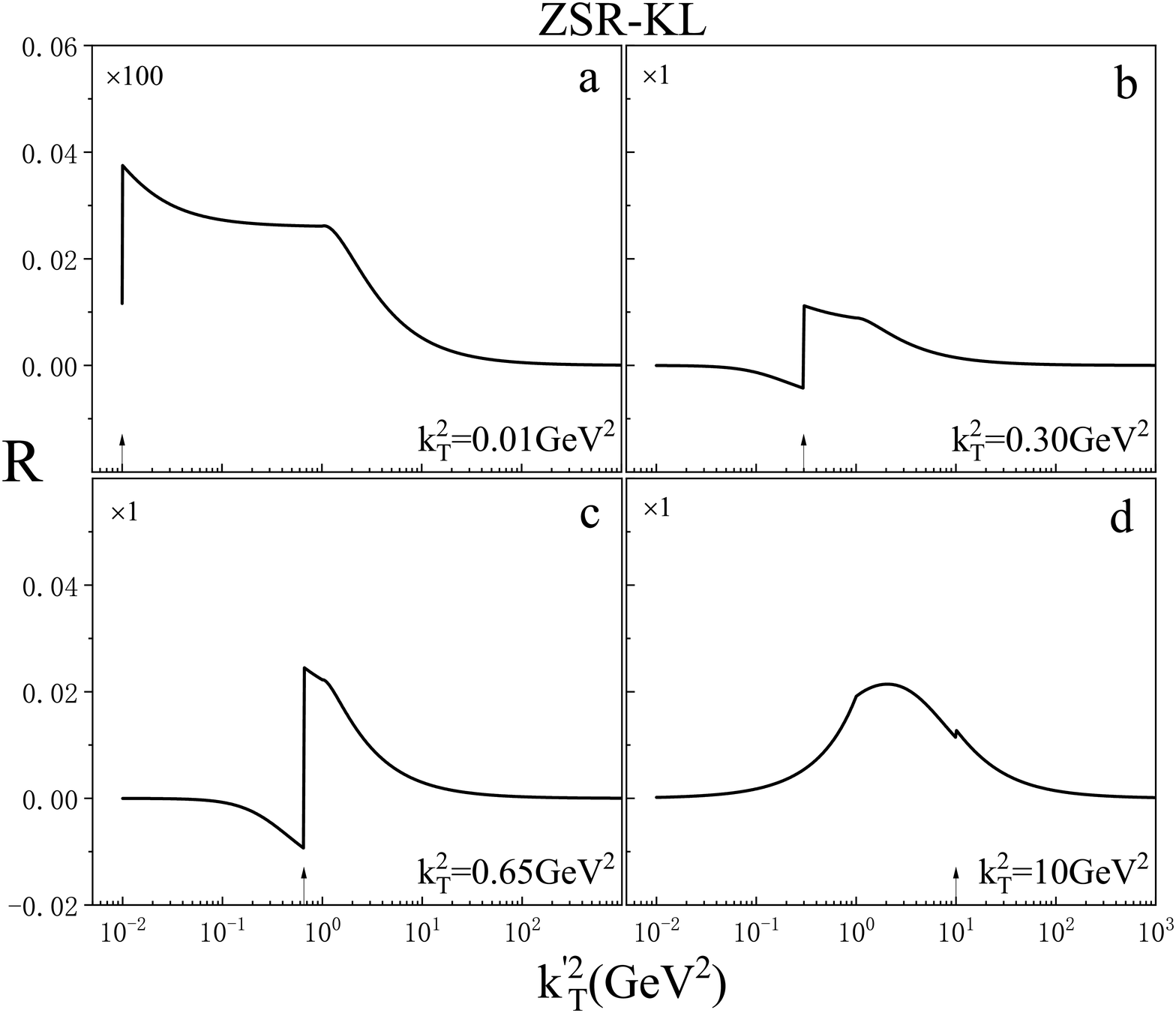} 
\end{center}
\end{figure}

\begin{figure}
\begin{center}
    \includegraphics[width=0.8\textwidth]{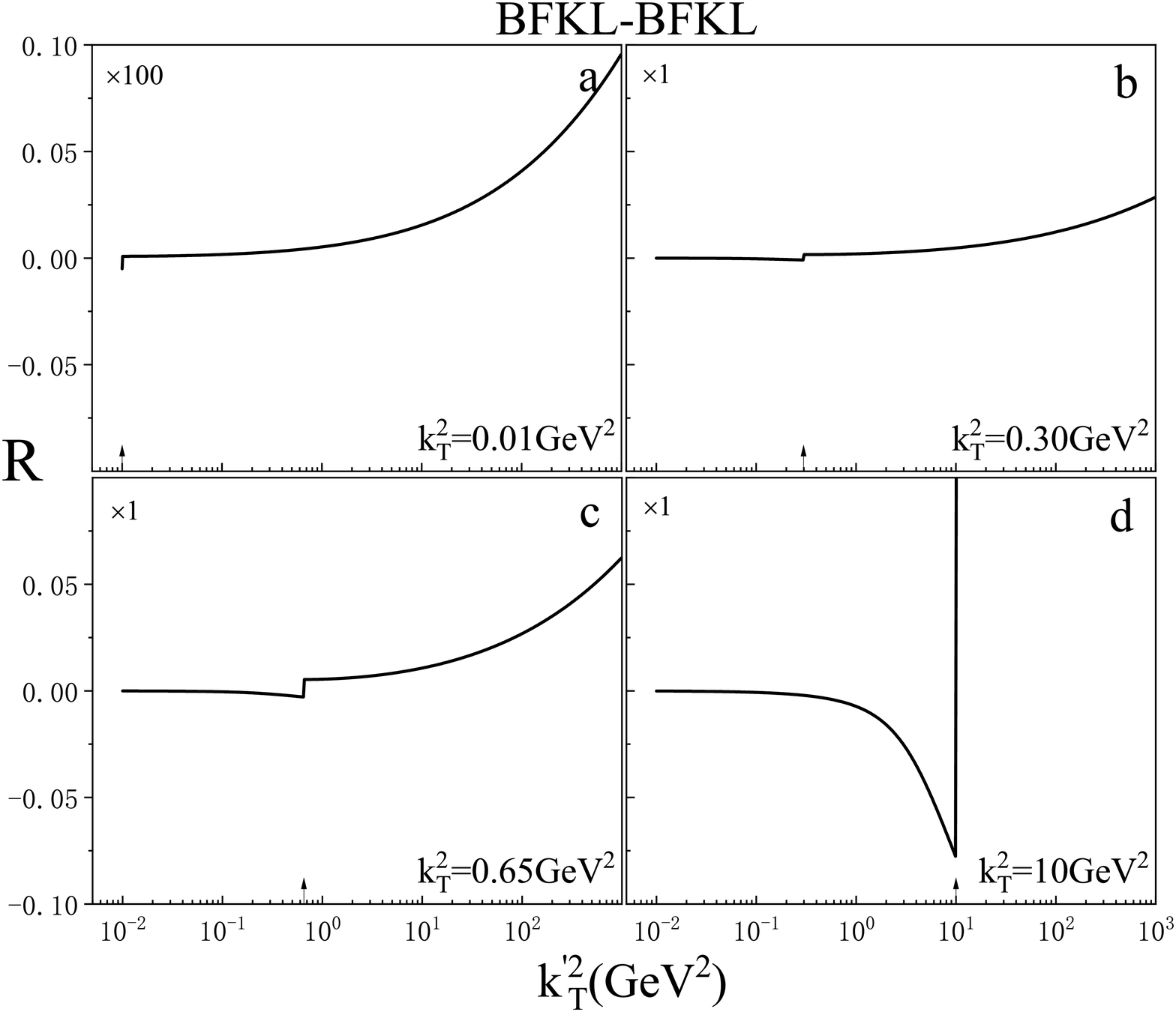} 
    \caption{The transverse momentum dependence of the integrated function $R$ in different evolution equations, where we use the GBW- and KL-inputs in Eq. (2.1)
and the BFKL-input in the BFKL equation. One can find the jump structure, which will arise the chaotic solutions in
Eq. (2.1).}\label{Fig.4}
\end{center}
\end{figure}

\begin{figure}
\begin{center}
    \includegraphics[width=0.8\textwidth]{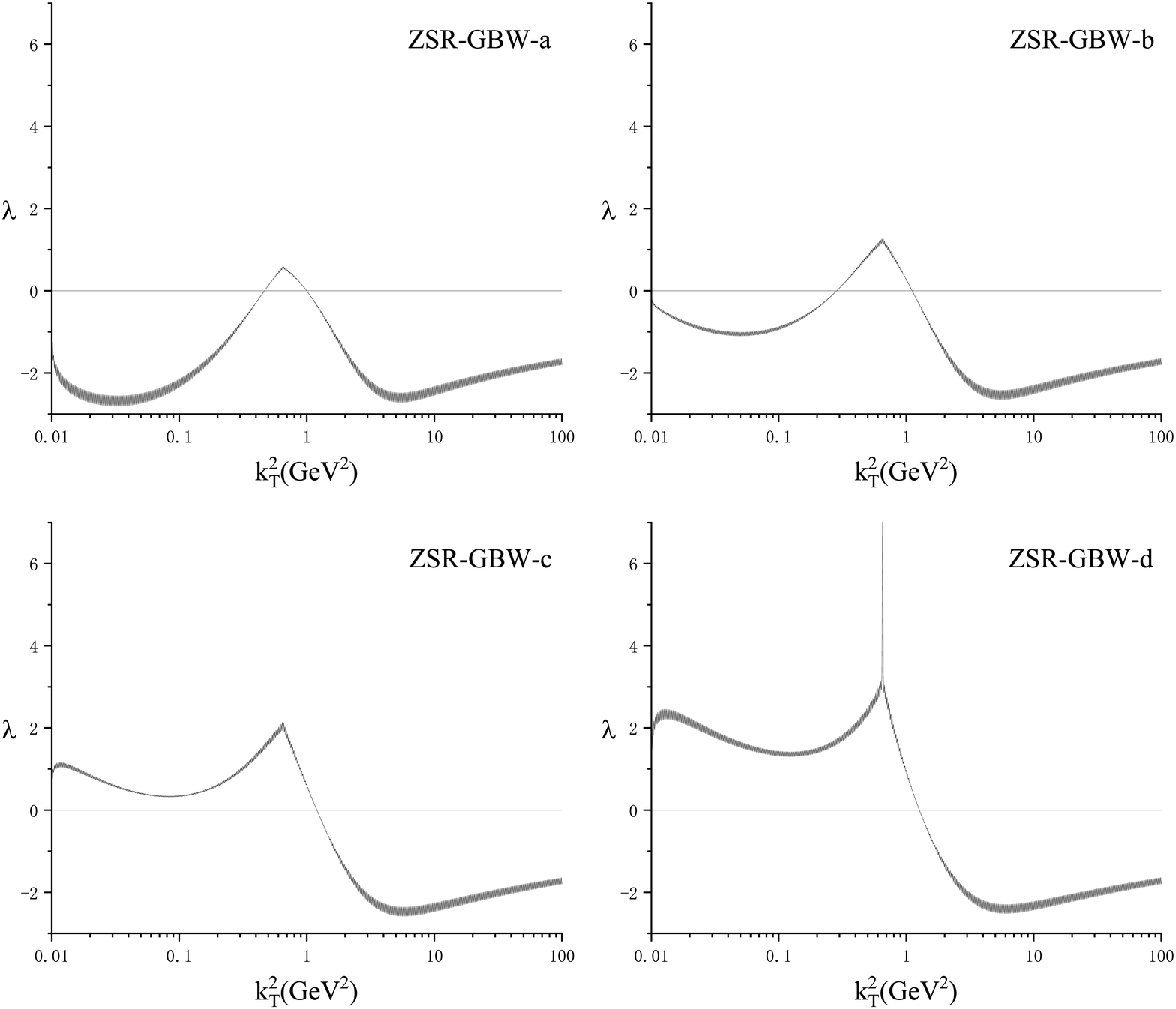} 
    \includegraphics[width=0.8\textwidth]{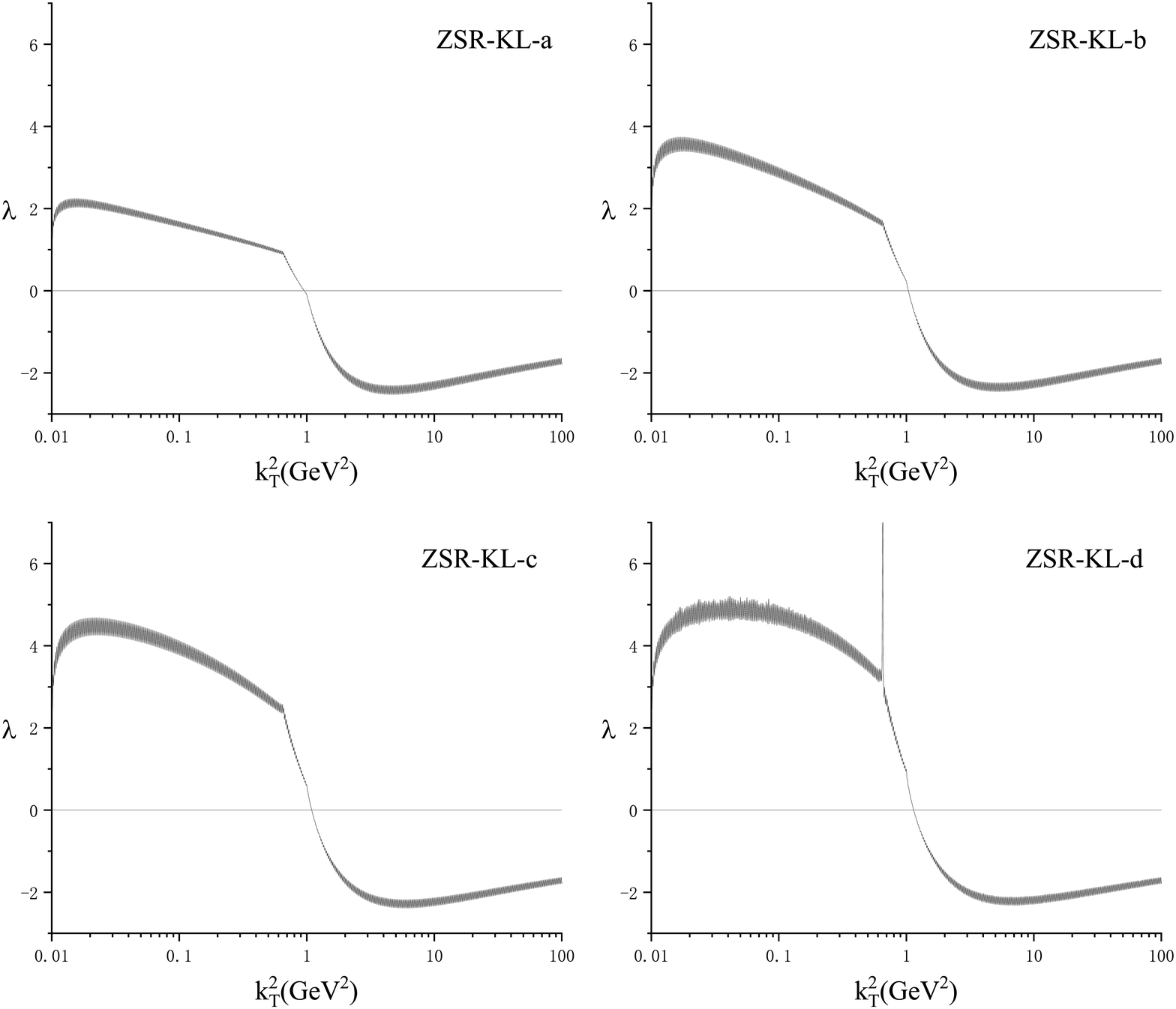} 
    \caption{The Lyapunov exponents of the chaotic solutions in Eq. (2.1) using the GBW and KL inputs.
   The standard program of the Lyapunov exponents see Ref. [5].
    }\label{Fig.5}
\end{center}
\end{figure}

    The critical momentum $k^2_c$ relates to the strength of the nonlinear terms of the QCD evolution equation. In order to verify
this property, we multiply
the nonlinear terms in Eq. (2.1) by a parameter $A$. Then we calculate the GC-critical momentum $k_c$ using different values
of $A$. The results in Table 1 show that the value of the critical momentum $k_c$
in a certain range is determined by the structure of the evolution equations. Using this result we can predict the nuclear target
dependence of the GC effect in the $p-A$ or $A-A$ collisions since the nonlinear terms of Eq. (2.1) are $A$-dependent. $A$ is the mass number of a nucleus.

\begin{table}
\caption{The relations of the GC-critic momentum $k_c$ with the nonlinear corrections.}
  \begin{center}
   \includegraphics[width=0.8\textwidth]{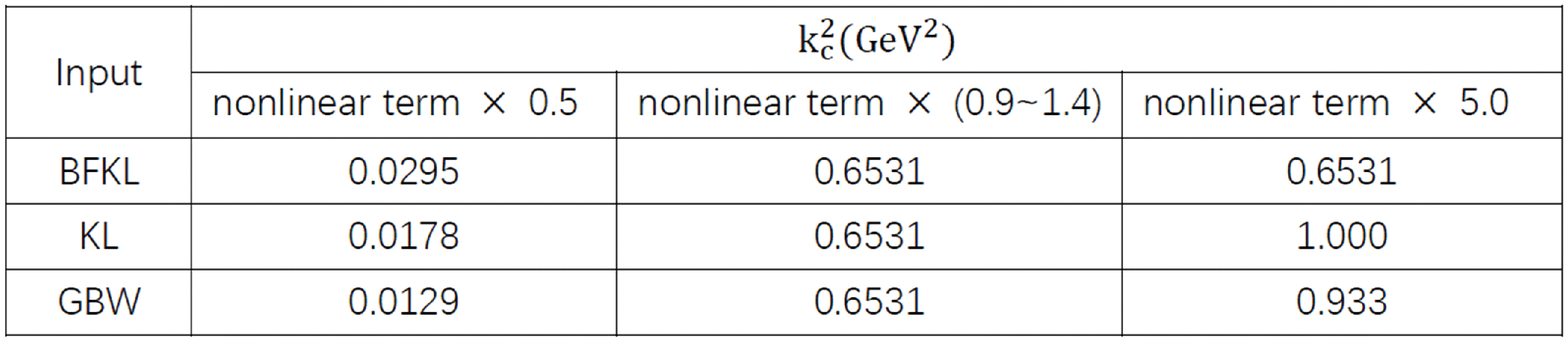}
    \end{center}
\end{table}

\begin{figure}
\begin{center}
    \includegraphics[width=0.8\textwidth]{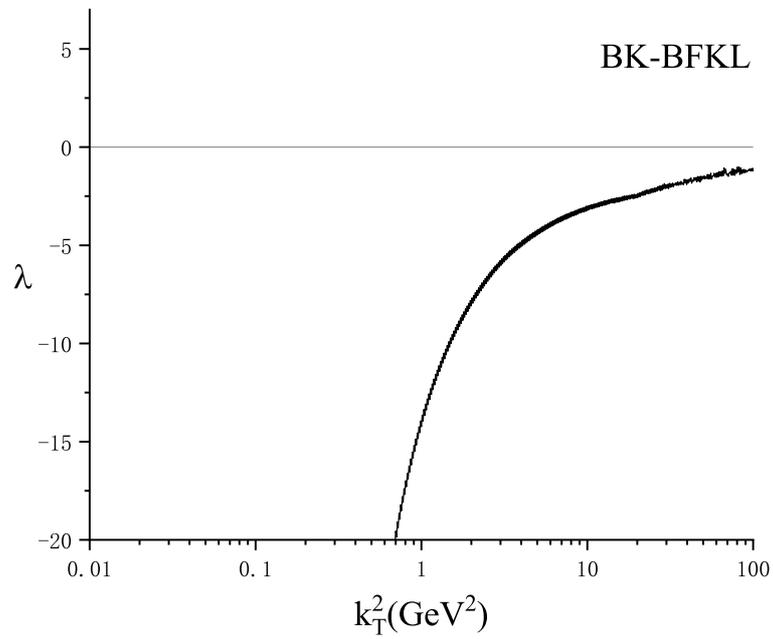} 
    \includegraphics[width=0.8\textwidth]{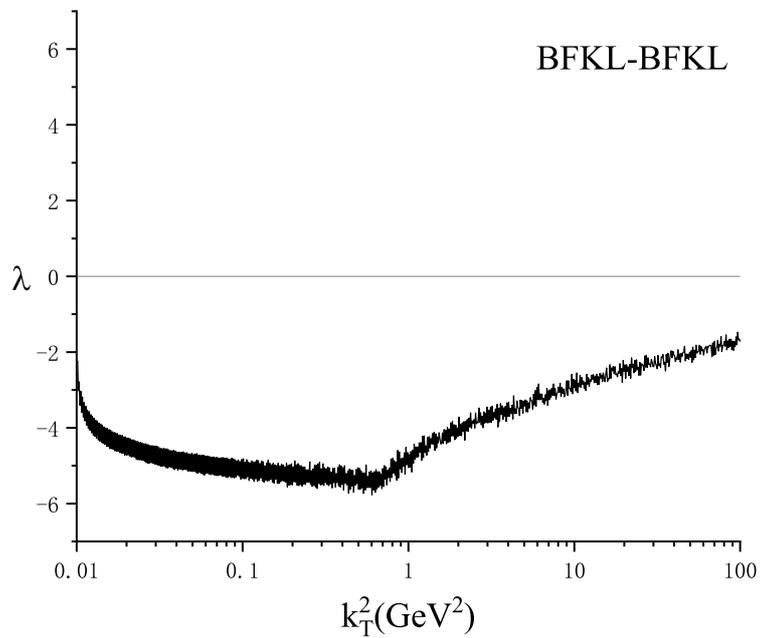} 
    \caption{The Lyapunov exponents of the BK and BFKL equations using the BFKL-input.}\label{Fig.6}
\end{center}
\end{figure}

     For comparison, we discuss the BK equation [10], which is generally considered as the typical
nonlinear corrections to the BFKL equation at the leading order
approximation. The BK equation can be written
in the full momentum space as [15,16]

$$-x\frac{\partial F(x,k_\perp)}{\partial x}$$
$$=\frac{\alpha_{s}N_c}{2\pi^2}\int d^2
k_\perp'
\frac{k_\perp^2}{(k_\perp-k_\perp')^2k_\perp^2}
2F(x,k_\perp')-\frac{\alpha_{s}N_c}{2\pi^2}
F(x,k_\perp)\int d^2 k'^2_\perp
\frac{k_\perp^2}{(k_\perp-k'_\perp)^2 k_\perp^2}$$
$$-\frac{18\alpha_s^2}{\pi R^2_N}\frac{N_c^2}{N_c^2-1}\frac{1}{k_\perp^2}
F^2(x,k^2_\perp), \eqno(2.5)$$ or taking the cylindrically
symmetric approximation

$$-x\frac{\partial F(x,k_\perp^2)}{\partial x}$$
$$=\frac{3\alpha_{s}k_\perp^2}{\pi}\int_{k'^2\perp}^{\infty} \frac{d k_\perp'^2}{k'^2_\perp}\left\{\frac{F(x,k'^2_\perp)-F(x,k_\perp^2)}
{\vert
k'^2_\perp-k_\perp^2\vert}+\frac{F(x,k_\perp^2)}{\sqrt{k_\perp^4+4k'^4_\perp}}\right\}
-\frac{81}{4}\frac{\alpha_s^2}{\pi
R^2_N}\frac{1}{k_\perp^2}F^2(x,k_\perp^2). \eqno(2.6)$$
We find that both the BK and BFKL equations
don't have the chaotic solutions since their Lyapunov exponents are negative (see Fig. 6).

     With only the chaotic effect it cannot produce the GC solutions in Eq. (2.1). The TOPT regularized nonlinear kernel plays its second
important role for converting the chaotic vibration into the strong shadowing and antishadowing effects, the later eventually forms the GC. Let us illustrate them. In Eq. (2.1) we have

$$\left[\frac{k_\perp^2F^2(x,k'^2_\perp)}{k'^2_\perp\vert
k'^2_\perp-k_\perp^2\vert}-\frac{k'^2_\perp F^2(x,k_\perp^2)}{k'^2_\perp\vert
k'^2_\perp-k_\perp^2\vert}\right]_{k'^2_\perp\sim k_\perp^2}\sim
\frac{d}{dk'^2_\perp}\left[\frac{k_\perp^2}{k'^2_\perp}
F^2(x,k'^2_\perp)\right]_{k'^2_\perp\sim k_\perp^2}. \eqno(2.7)$$
Once chaos is produced, the fast oscillations of the gluon density will generate both the negative
and positive nonlinear corrections to the increment $\Delta F(x,k^2_\perp)$ through
derivative operations. Usually we call the negative or positive corrections to the nonlinear
evolution equations as the shadowing or antishadowing effects.
The former gradually suppresses the grownup of
the gluon density since $F^2$ is also suppressed, while the later is a positive feedback process and it increases rapidly
due to it is proportional to the increasing $F^2$. Let us consider Fig. 7, which
presents the gluon distributions when the nonlinear terms in Eq. (2.1) take positive absolute values
(curve 1), negative absolute values (curve 2) and zero value (dashed curve), they corresponding to
the pure antishadowing effect, the pure shadowing effect and the linear BFKL evolution, respectively.
One can find that the pure shadowing effect gradually suppresses the increasing gluon distribution, while the pure antishadowing leads to a fast divergence of the gluon density.

\begin{figure}
    \begin{center}
        \includegraphics[width=0.8\textwidth]{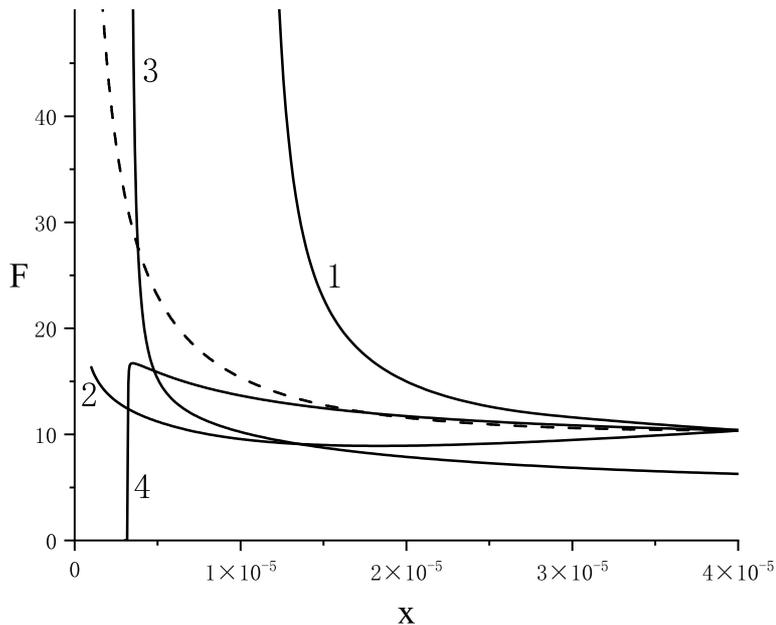}
        \caption{The net antishadowing effect (curve 1) and net shadowing effect (curve 2). Dashed curve is a solution
of the linear BFKL equation. A pair of curves 3 and 4 are the solutions of Eq (2.1), where the shadowing and antishadowing are correlated by momentum conversion and they lead to the GC.}\label{Fig.7}
    \end{center}
\end{figure}

    Note that the above examples show the isolate shadowing and antishadowing effects.
In fact, the shadowing and antishadowing are correlated due to the local momentum conservation.
Although Eq. (2.1) works at the small-$x$ range and it does not describe the total momentum conservation,
the local momentum conservation is still valid at every QCD vertex. The TOPT cutting rule in the derivation of Eq. (2.1)
emphasises to sum all possible Feynman diagrams including real and virtual processes at the same order approximation,
as they naturally hold the local momentum conservation [5].
It leads to the correlation between the negative shadowing effect and the positive antishadowing effect, since these two effects
origin from virtual and real diagrams, respectively. Thus, a strong antishadowing effect in Eq. (2.1) $must$ be
accompanied by the disappearance of the shadowed gluons due to the local momentum conservation as shown by a pair of curves 3 and 4 in Fig. 7. A similar example for the correlation of shadowing and antishadowing appears in a modified DGLAP equation [4],
where the antishadowing effect is
weaker than the shadowing effect since the integral range is restricted and the net
shadowing effect dominates the process. The restriction of the resulting antishadowing effect has been confirmed
by the observed EMC effect [17]. Thus, we do not need to design a mysterious mechanism to condense
the gluons, the momentum conservation plus the net antishadowing effect in Eq. (2.1) may realize the GC.

      Therefore, the GC is the combined effect of chaos and antishadowing in Eq. (2.1), which
forms a GC-chain: the potential perturbation arises
chaotic oscillations in the BFKL dynamics $\rightarrow$generating the positive antishadowing corrections$\rightarrow$converting the shadowed gluons to the gluons with the critical momentum $(x_c,k_c)$. Therefore, the gluon condensation
is a natural consequence of random evolution of gluons at very high energy hadronic processes.

    The unintegrated gluon distribution $F(x,k_{\perp}^2)$ of a bound nucleon in the nucleus $A$ satisfies
the following evolution equation

$$-x\frac{\partial F(x,k_{\perp}^2)}{\partial x}$$
$$\simeq\frac{3\alpha_{s}k_{\perp}^2}{\pi}\int_{k^2_{\perp,min}}^{\infty} \frac{d
k'^2_{\perp}}{k'^2_{\perp}}\left\{\frac{F(x,k'^2_{\perp})-F(x,k_{\perp}^2)} {\vert
k'^2_{\perp}-k_{\perp}^2\vert}+\frac{F(x,k_{\perp}^2)}{\sqrt{k_{\perp}^4+4k'^4_{\perp}}}\right\}$$
$$-\frac{81}{16}\frac{\alpha_s^2A^{1/3}}{\pi R^2_N}\int_{k^2_{\perp,min}}^{\infty} \frac{d
k'^2_{\perp}}{k'^2_{\perp}}\left\{\frac{k_{\perp}^2F^2(x,k'^2_{\perp})-k'^2_{\perp}F^2(x,k_{\perp}^2)}
{k'^2_{\perp}\vert
k'^2_{\perp}-k_{\perp}^2\vert}+\frac{F^2(x,k_{\perp}^2)}{\sqrt{k_{\perp}^4+4k'^4_{\perp}}}\right\}.
\eqno(2.8)$$ We take the K-L model in a nucleus $A$ as the input of Eq. (2.8) at $x_A$

$$F(x_A,k_{\perp}^2)=\left\{
\begin{array}{ll}
f_0k^2_{\perp} ~~~~if~k^2_{\perp}\leq Q_{A,s}^2\\\\
f_0Q^2_{A,s}~~~~if~k^2_{\perp}>Q_{A,s}^2 \\
\end{array} \right.,\eqno(2.9)$$where $Q^2_{A,s}=Q^2_sA^{1/3}$, $Q_s=1~GeV $ is the saturation scale of a free nucleon [18].

\begin{figure}
	\begin{center}
		\includegraphics[width=0.8\textwidth]{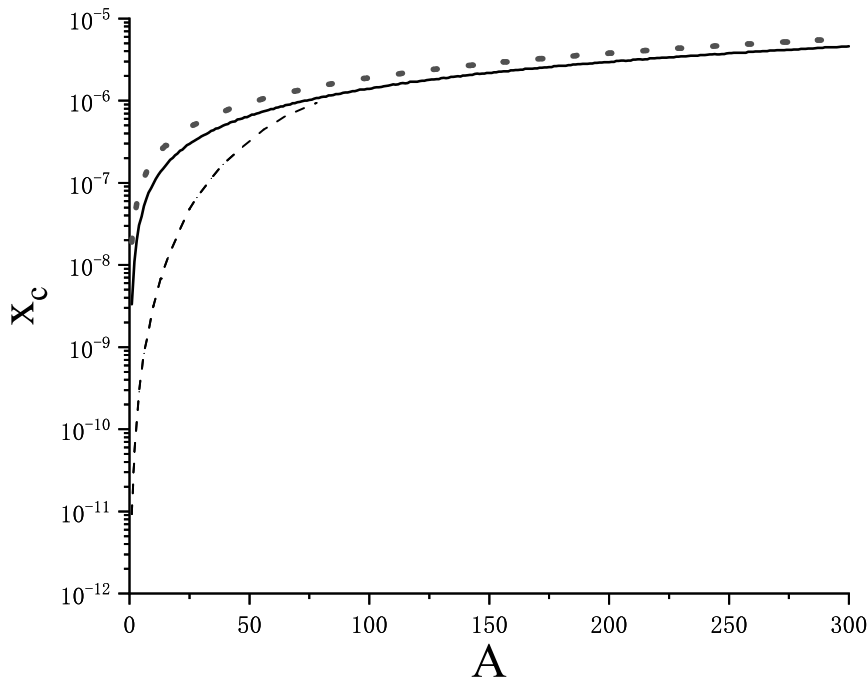}
		\caption{Predicted critic momentum $x_c$ (solid curve) in different nuclear target. The point curve is
$x_A=1.9\times 10^{-8}A$. The dashed curve is the possible modified $x_c$ due to Eq. (3.16).
}\label{Fig:8}
	\end{center}
\end{figure}

\begin{figure}
	\begin{center}
		\includegraphics[width=0.8\textwidth]{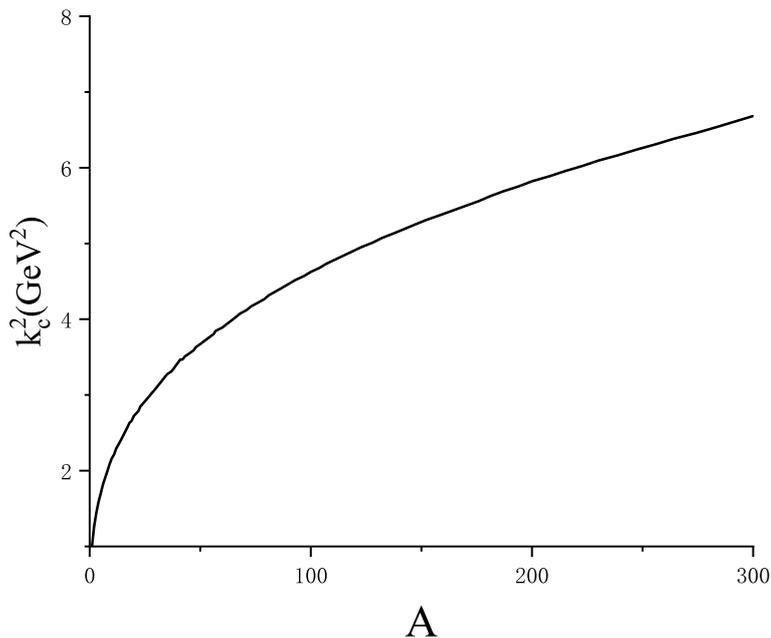}
		\caption{Predicted critic momentum $k^2_c$ in different nuclear targets.
		}\label{Fig:9}
	\end{center}
\end{figure}

    Generally, the bigger the nucleus, the earlier Eq. (2.8) works.
The GC-effect originates from the chaotic solution of Eq. (2.8).
Once a most robust chaos is created, it will dominate the whole process of the condensation, no matter
the event occurs whether in the longitudinal $(\sim A^{1/3})$ or horizontal $(\sim A^{2/3})$ area of the nucleus.
Therefore, the starting point of evolution in Eq.(2.8) is proportional to the volume of a nucleus rather than its
longitudinal scale, i.e., we take $x_A=x_pA$, $x_p$ is the starting point of the Eq. (2.8) in a free nucleon.
We use the input Eq. (2.9) and $x_p\equiv1.9\times 10^{-8}$ to calculate the relation $(x_c,k_c)\sim A$.  The results are drawn
in Figs. 8 and 9. We will use them to discuss the GC effect in hadron collisions in the following sections.

\section{The GC-effect in the planning hadron collides beyond the LHC energies}

     We abstract some formulas for the applications of the GC effect in high energy hadron-hadron collisions and
the detailed derivations can be found in works [6-9].  According to QCD, the yield of secondary
particles at the high energy $p-p$
collision relates to the number of excited gluons, which participate in the
multi-interactions. The number of pion will rapidly grow when a lot of gluons
enter the interaction range due to the GC effect. Without concrete
calculations, one can image that this will form an excess phenomenon
in the cosmic-ray spectra. We focus on the central rapidity region, where
the multi-hadrons (mainly pions) are produced due to the excited gluons. This process contains two steps

$$p+p\rightarrow gluons~ (or~gluons~and~quark-antiquark~pairs)\rightarrow n_{\pi}, \eqno(3.1)$$$n_{\pi}$ is the multiplicity
of the secondary pions.

   Let us deeply understand the hadronization processes in the QCD-view.
Based on the Dyson-Schwinger (DS) equation, the recent research by Roberts, et al., shows that the effective quark mass in hadron might be dynamically generated due to the gluon nonlinearities [19].
In this mechanism the nonperturbative gluon propagator is constructed by a lot of partons (i.e., gluons and sea quarks, note that
the later is originated from gluons) in the hadron infinite momentum frame. A
massless quark has got a small current quark mass through the Higgs mechanism and it further
acquires the effective mass through the
nonperturbative gluon propagator in the QCD self-energy diagram (Fig. 10). These "dressed" quarks are called as the constituent quarks.
Now we compare the gluon distributions of the gluon condensate
and the glasma. At the same collision energy $\sqrt{s}$,
the gluon condensate has abundant gluons to greatly generate the number of the constituent quarks and increase the number $n_{\pi}$,
i.e., $n_{\pi,GC}\gg n_{\pi,Glasma}$.  On the other hand, we have $1/2\sqrt{s}\geq n_{\pi}m_{\pi}$ due to the restriction of the energy conservation. As a limit, we image that almost all available energy of the collision at the center-of-mass system is used to create pions.

\begin{figure}
    \begin{center}
        \includegraphics[width=0.8\textwidth]{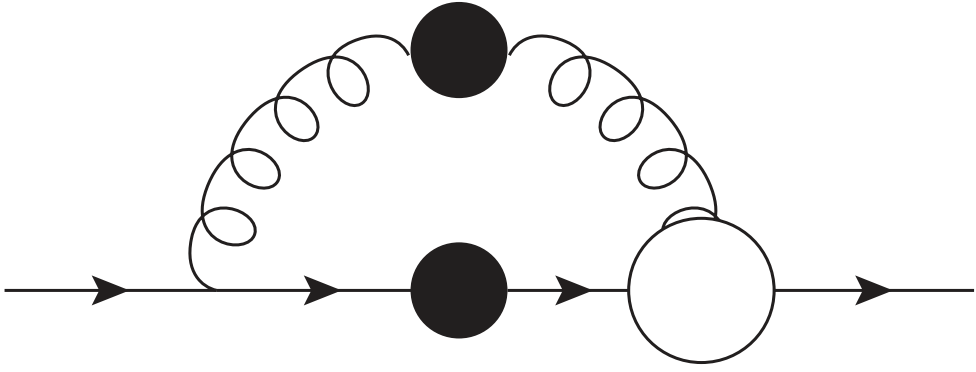}
        \caption{The self-energy diagram of obtaining effective constituent quark mass. Works [19] use it to explain the formation of
pion mass. Empty circle is the DS vertex
and black spots indicate the nonperturbative propagators, which contain the multi-parton components.}\label{Fig.10}
    \end{center}
\end{figure}

    Combining the relativistic invariant and energy conservation, one
straightly writes [6,7]

$$(2m_p^2+2E_pm_p)^{1/2}=E^*_{p1}+E^*_{p2}+n_{\pi}m_{\pi}, \eqno(3.2)$$
$$E_p+m_p=m_p\gamma_1+m_p\gamma_2+n_{\pi}m_{\pi}\gamma, \eqno(3.3)$$where
$E^*_{pi}$ is the energy of leading proton at the center-of-mass system,
$\gamma_i$ is the corresponding Lorentz factor.
We can easily get the solutions of $n_{\pi}(E_p,E_{\pi})$ in
the $p-p$ collision

$$\ln n_{\pi}=0.5\ln E_p+a, ~~\ln n_{\pi}=\ln E_{\pi}+b,\eqno(3.4)$$  where $E_{\pi}
\in [E_{\pi}^{GC},E_{\pi}^{max}]$.  The parameters

$$a\equiv 0.5\ln (2m_p)-\ln m_{\pi}+\ln 1/2, \eqno(3.5)$$ and
$$b\equiv \ln (2m_p)-2\ln m_{\pi}+\ln 1/2. \eqno(3.6)$$ These equations give
the one-to-one relation among $n_{\pi}$, $E_p$ and
$E_{\pi}^{GC}$. Equation (3.4) gives the following special relation

$$E_p=e^{-2(a-b)}E_{\pi}^2=\frac{2m_p}{m_{\pi}^2}E_{\pi}^2,  \eqno (3.7)$$

    The upper limit $E_{\pi}^{max}=14(E_{\pi}^{GC})^2$ (in the GeV-unit) if the charged
particles can be accelerated continuously. However the actual $E_{\pi}$ cannot reach such high limit
because the restriction of acceleration mechanism. Therefore, we use a smaller cut-energy $E_{\pi}^{cut}$ to replace
$E_{\pi}^{max}$. $E_{\pi}^{cut}$ relates to the accelerator mechanism and $E_{\pi}^{cut}\gg E_{\pi}^{GC}$.
   According to the numerical simulation, $E_{\pi}^{cut}$ is required at least one order of magnitude greater than $E_{\pi}^{GC}$, i.e., $E_{\pi}^{cut}\geq 10E_{\pi}^{GC}$.
$E_{\pi}^{GC}$ and $(x_c,k_c)$ have the following relation [6,7]

$$E_{\pi}^{GC}=\exp\left(0.5\ln\frac{k^2_c}{2m_px_c}+a-b\right)=\frac{m_{\pi}}{2m_p}\frac{k_c}{\sqrt{x_c}}.\eqno(3.8)$$Using this equation,
one can predict the GC-threshold in Fig. 11.
We find
that the distribution of $E_{\pi}^{GC}$ can be roughly divided into three ranges: (i) $E_{\pi}^{GC}\approx 100~GeV$
for intermediate and heavy nuclei $(A>100)$, (ii) $E_{\pi}^{GC}$ quickly increases for light nuclei ($A<20$) and (iii) the area between them $(20<A<100)$.

\begin{figure}
	\begin{center}
		\includegraphics[width=0.8\textwidth]{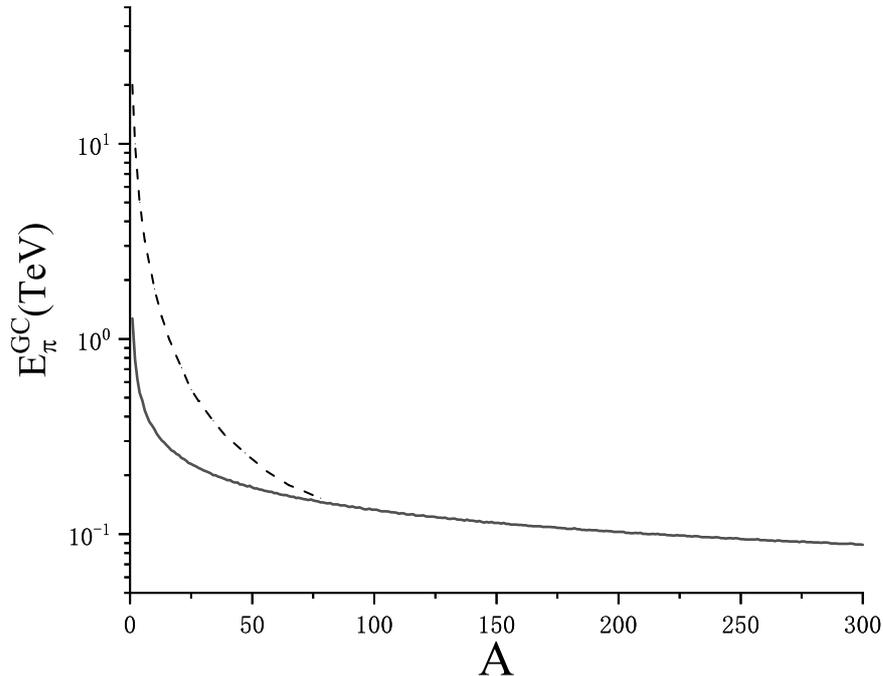}
		\caption{The predicted GC-threshold $E_{\pi}^{GC}$ in the $p-A$ (or $A-A$) collisions (solid curve), which are
obtained by Eq. (3.8) and the data of Figs. 8 and 9. The dashed curve is the possible modified GC-threshold $E_{\pi}^{GC}$ due to
the corrections of Eq. (3.8).
		}\label{Fig:11}
	\end{center}
\end{figure}

    The contributions of proton and nucleus in the $p-A$ collisions dominate the
rapidity distribution on the two sides of rapidity space, respectively.
Note that the GC effect begins work when the gluons with $x_A$ takes the first to participate the $A'-A$ collisions if $A>A'$,
and enhances the total cross section of the collisions,
therefore, we have $E_{\pi}^{GC}(p-A)\simeq E_{\pi}^{GC}(A-A)$, although the GC-signal strength of the former is half weaker than the latter.

\begin{table}[htbp]
    \caption{The predicted center-of-mass energy $\sqrt{s_{pA}^{cut}}$ appearing the GC effect and the relating
    quantities in different hadronic collisions (in the GeV-unit).
    }\label{tab:2}
    \vskip 0.1cm
    \includegraphics[width=1.05\columnwidth,angle=0]{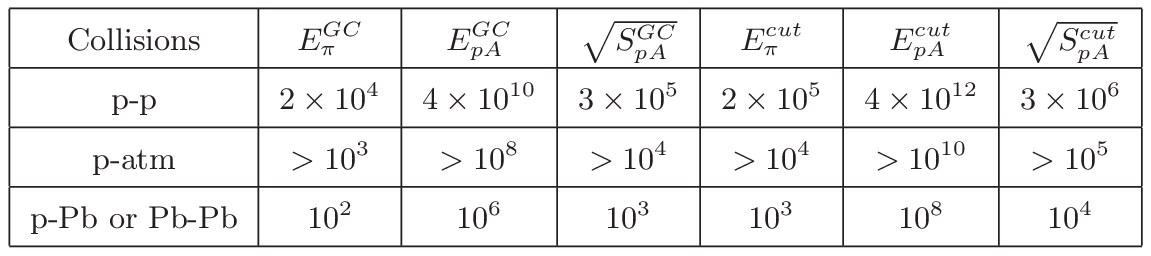}
\end{table}

   As we have known, the GC effect causes a big excess in the cross section
of hadron collisions at center-of-mass energy $\sqrt{s^{GC}}$. In Tab. 2 we give these thresholds using Eqs. (3.7), (3.8)
and the modified curve is shown in Fig. 11.
For example, we take $E_{\pi}^{GC}(p-p)=20$ TeV, thus, $E_{pp}^{GC}=100\times (2\times 10^4)^2$ GeV$=
4\times 10^{10}$ GeV and $\sqrt{s_{pp}^{GC}}=\sqrt{2m_pE_{pp}^{GC}}=
3\times 10^2$ TeV. To observe the GC effect in the $p-p$ collision, one needs $E_{\pi}^{cut}\geq 10E_{\pi}^{GC}$, which requires  $\sqrt{s_{pp}^{cut}}\geq 3\times 10^3$ TeV.
The maximum energies of the $p-p$ collision at the LHC is $\sqrt{s_{pp}}=13$ TeV $\ll 3\times 10^3$ TeV.
Obviously, we can not record any GC-signals from the $p-p$ collision in the recent LHC.

    Now we consider the the $p-Pb$ and $Pb-Pb$ collisions at the LHC, their maximum energies are
8.16 TeV and 5.02 TeV, respectively. Referring to Fig. 11, we take $E_{\pi}^{GC}(Pb-Pb)=100$ GeV for them.
The results are presented in Tab. 2. The energy region where the GC-signal appears is $\sqrt{s_{pPb}^{cut}}>10$ TeV.
Although the energies of these collisions in the LHC are already close to the GC-thresholds,
it is still necessary to further increase the collision energy.

    We emphasize that the GC effect may efficiently convert the kinetic energy of the parent protons
into a large number of secondary
particles. In fact, the multiplicity of mini gluon-jets in the hadronic collisions with the GC effect is
about $10^3-10^4$ times larger than that without the GC model [5].
Besides, there is a highest gluon-pion conversion rate in the GC model.
Thus, about half of the kinetic energy of protons are converted
to the large numbers of photons with energy $E_{\gamma}=m_{\pi}/2$ in the center of mass (CM) frame in the $Pb-Pb$ collision. Such monochrome gamma-rays have extra high strength
in a narrow space, and may damage the detectors in the laboratory.

    We noticed that the Auger collaboration indirectly used the cosmic ray data at the top of the atmosphere and found
that $pA$ cross section at $\sqrt{s}\sim 100~TeV$ is the
normal value $\sim 567~mb$ with no big increment [21]. According to Fig. 12 we
guess that the GC-threshold $E_{\pi}^{GC}(p-atm)> 1~TeV$ for light nuclei at the top of atmosphere.
which implies $\sqrt{s_{p-atm}^{GC}}> 10~TeV$, and $\sqrt{s_{p-atm}^{cut}}> 100~TeV$. This result means
that the observed energy by Auger collaboration at $\sqrt{s}=100~TeV$ is close to the
lower-limit of the GC-effect range if the above mentioned indirect estimations are correct.

   The more direct method of probing the GC effect in the laboratory is
the Electron Ion Collider (EIC) [22] and the Large
Hadron electron Collider (LHeC) [23].
The designed collision energy of upcoming EIC(US) is
$\sqrt{s_{eA}}\simeq 140~GeV$, which is in the CGC range ($x>10^{-5}, Q^2>1~GeV^2)$. According to our estimation in Fig.8, the range
of the GC effect is $(x_c\simeq 10^{-6},  k^2_c\simeq 5~GeV^2)$, which exceeds the range of the EIC(US).
However, the center-of-mass energy of the proposed Large Hadron electron Collider (LHeC) $\sqrt{s_{eA}}\sim 1$ TeV
will be able to cover a very low $x$-range: $x\sim 10^{-6}$ at $Q^2>1 GeV^2$ in $eA$ collisions, where we may
record the GC-signal.

    Since the huge number of gluons are condensed in a critical momentum $(x_c,k_c)$,
it should greatly increase the hadron cross section and release strong gamma-rays.
We warn that further increase of the hadron collision energies in the next LHC plans may lead to unexpected intense gamma-rays in the accelerator, they look like the artificial mini GRBs and may damage the detectors

\section{Discussions}

    (1) The evolution of the BFKL dynamics may
become nonperturbative near the singular range mentioned above.
Fortunately, both the lattice simulations and the nonperturbative dynamics of the QCD
show that the effective strong coupling constant is restricted by
$\alpha_s/\pi\leq B$ ($B$ is a constant) [24].

    (2) The equation (2.1) is based on the leading order approximation, where
the higher order corrections are neglected. An important question
is: will the chaos effects disappear in the evolution equation after
considering higher order corrections? Works [5] have discussed this problem and we abstract some conclusions as follows.
The GC effect origins from the singular nonlinear
evolution kernel and local momentum conservation, they are general structure at the leading
$\ln(1/x)$-resummation. The multi-singular structure from higher order
corrections should be cancelled by the contributions of the virtual Feynman diagrams at a same order level.
The resulting nonlinear evolution equation is still keeping the conditions of existing GC.
Although the higher order QCD corrections may change the value of $k_c$, however, the simple form of the GC solution contains only a few parameters $(x_c,k_c)$ and they can be
determined by the experimental data.

    (3) The BK equation is generally considered as a typical
nonlinear correction to the BFKL equation at the $LL(1/x)$
approximation and it is the most widely used small $x$ evolution in the community.
Why the BK equation doesn't have the chaotic solution? We try to answer this question.
The BK equation is usually written by using the scattering
amplitude $N(x_\perp, x)$ in the transverse coordinator space

$$-x\frac{\partial N(x_{\perp},x)}{\partial x}$$
$$=\frac{\alpha_{s}N_c}{2\pi^2}\int d^2 x'_{\perp}
\frac{x_{\perp}^2}{x_{\perp}^2(x_{\perp}-x'_{\perp})^2}\left[N(x'_{\perp},x)+N(x_{\perp}-x'_{\perp},x)-N(x_{\perp},x)\right.$$
$$-N(x'_{\perp},x)N(x_{\perp}-x'_{\perp},x)].\eqno(4.1)$$ The linear part corresponds to the BFKL evolution equation, while
the nonlinear evolution kernel is regularized by
the connecting amplitude $N(x'_{\perp},x)N(x_{\perp}-x'_{\perp},x)$ in the dipole splitting model to avoid the Lipatov-singularity.
In order to illustrate the physical meaning of this regularization method, we
use

$$N(x,k_\perp)=\int
\frac{d^2x_\perp}{2\pi}\exp(-ik_\perp\cdot x_\perp)
\frac{N(x_\perp,x)}{x_\perp^2},\eqno(4.2)$$to rewrite the BK equation in the transverse momentum space

$$-x\frac{\partial N(x,k_\perp^2)}{\partial x}$$
$$=\frac{3\alpha_{s}}{\pi}\int_{k'^2\perp}^{\infty} \frac{d k'^2_{\perp}}{k'^2_\perp}\left\{\frac{k'^2_{\perp}N(x,k'^2_{\perp})
-k^2_{\perp}N(x,k_\perp^2)}
{\vert
k'^2_\perp-k_{\perp}^2\vert}+\frac{k^2_{\perp}N(x,k_\perp^2)}{\sqrt{k_\perp^4+4k'^4_\perp}}\right\}
-\frac{3\alpha_s}{\pi}N^2(x,k_\perp^2). \eqno(4.3)$$ If we define

$$N(x,k_\perp)\equiv
\frac{27\alpha_s}{16k_\perp^2R^2_N}F(x,k_\perp)
,\eqno(4.4)$$one can obtain Eq. (2.6) for the unintegrated gluon distribution $F(x,k_{\perp})$.

    Let us to expose a relation between the ZSR and BK equations.
The Lipatov-singularity in the nonlinear evolution kernel is arisen
by the crossed diagrams of two different amplitudes in Fig. 1d and they imply the random evolution in the transverse
momentum space. A simple, but unreasonable way for avoiding the singularity is to remove these crossed diagrams. In this case,
one can find
that Fig. 1d reduces to Fig. 1c. The latter is the contributions of the gluon fusion to the DGLAP evolution equation at
the $DLL$-approximation, i.e.,

    $$Q^2\frac{\partial G(x,Q^2)}{\partial Q^2}=-\frac{36\alpha_s^2}{8Q^2R^2_N}\frac{N_c^2}{N_c^2-1}\int\frac{dx'}{x'}G^2(x',Q^2),\eqno(4.5)$$where the
transverse momenta of gluons are strongly ordered [3,4]. Note that
the relation between gluon distribution $G(x,Q^2)$ and unintegrated gluon distribution $F(x,k^2_{\perp})$ is

$$G(x,Q^2)\equiv \int^{Q^2}_{k^2_{\perp,min}}\frac{dk^2_{\perp}}{k^2_{\perp}}F(x,k^2_{\perp}).  \eqno(4.6)$$

    We use $F(x,k^2_{\perp})$ to replace $G(x,Q^2)$. From Eq.(5.5) we have

    $$\Delta G(x,Q^2)=-18\alpha_s^2\frac{N^2_c}{N_c^2-1}\int^{Q^2}_{Q^2_{min}}\frac{dk^2_{\perp}}{k^2_{\perp}}\int\frac{dx'}{x'}
F^{(2)}(x',k^2_{\perp}), \eqno(4.7)$$or

$$\Delta F(x, k^2_{\perp})=\left.Q^2\frac{\partial \Delta G(x,Q^2)}{\partial Q^2}\right \vert_{Q^2=k^2_{\perp}}
=-\frac{18\alpha_s^2}{\pi k^2_{\perp}R^2_N}\frac{N^2_c}{N^2_c-1}\int\frac{dx'}{x'} F^2(x',k^2_{\perp}), \eqno(4.8)$$and

$$-x\frac{\partial F(x, k^2_{\perp})}{\partial x}
=-\frac{18\alpha_s^2}{\pi k^2_{\perp}R^2_N}\frac{N^2_c}{N^2_c-1}F^2(x,k^2_{\perp}), \eqno(4.9)$$Where
we define

$$F^{(2)}(x,k^2_{\perp})\equiv\frac{1}{\pi^2R_N^2}F^2(x,k^2_{\perp}). \eqno(4.10)$$Combining Eq. (4.9) with the linear BFKL equation,
we can obtain the BK equation (2.6).

    This example exposes a fact: the nonlinear part of the BK
equation neglects the contributions of the random evolution of gluons on the transverse space.
Therefore, the BK equation losses the GC-source since it
avoids the Lipatov-singularity in the nonlinear kernel by using a dipole model.
Thus, the BK equation and its general form [13] have not the GC solution.

    Interestingly, we show the following evolution of the QCD evolution equations with increasing gluon densities:

   $$ The~ DGLAP~ equation~(\vert a\vert^2)~[Violation~ of ~ unitarity]$$
   $$\rightarrow ~The~ BFKL~ equation~(\vert b\vert^2)~[Violation~of~ unitarity]$$
   $$\rightarrow~ The~ GLR-MQ-ZRS ~equation~(\vert a\vert^2+\vert c\vert^2)~[Shadowing~ and~ antishadowing]$$
   $$\rightarrow~ The~ BK equation~(\vert a\vert^2+\vert c\vert^2)~[Saturation]$$
   $$\rightarrow~ The~ ZSR~ equation~(\vert b\vert^2+\vert d\vert^2)~[Gluon~ condensation],$$where the brackets mark the
amplitude structure using Fig. 1, and
square brackets contain the predicted effects. This picture reflects the self-consistence of the evolution equations
including Eq. (2.1).

    In summary, we present our new understanding of the GC effect in a
nonlinear QCD evolution equation. Through the numerical solutions
of this evolution equation, we find that the random evolution of gluons in the
traverse momentum space generates the Lipatov-singularity both in the linear and nonlinear terms.
After regularization of these singularities according to the standard quantum field theory, the equation emerges
a special nonlinear structure, which causes the chaotic oscillations and then generates the strong net antishadowing effect.
Due to the restriction of the local momentum conservation, the antishadowing effect gathers the shadowed gluons to a state at a critical momentum, and forms the gluon condensate. We estimate the parameters in the GC solution, which predict
that the hadron collisions $p-Pb$ and $Pb-Pb$ in the LHC are already close to
the energy region of the GC effect. Since the huge number of gluons are condensed in a critical momentum,
it should greatly increase the hadron cross section and release strong gamma-rays.
We warn that further increase of the hadron collision energies in the next LHC plans may lead to unexpectedly intense gamma-rays in the accelerator, they look like the artificial mini GRBs and may damage the detectors.

{\it Acknowledgments:} This work is supported by the National
Natural Science of China (No.11851303). Q.H. Chen acknowledges
support from the National Natural Science of China (No.12147208).

\end{document}